\documentclass{article}
\usepackage[dvips]{graphicx}
\usepackage{epsfig}
\usepackage{longtable}
\usepackage{float}
\title{    Maxwell viscoelastic dynamics  of the DNA in the THz range.}
\author{ V.L.Golo}

\begin{document}

\maketitle

\begin{abstract}
  The attenuation of phonon modes of the DNA is due to the exchange of 
  water molecules  adsorbed by a molecule of DNA and the bulk solvent. Using Maxwell's mechanism
  of relaxation and a simple lattice model of the DNA, we show that the attenuation tends to a
  constant value for  phonon frequencies larger than the inverse residence time of water molecules. 
  We come to the conclusion that in the THz range 
  the attenuation could be small enough to allow the propagation of the phonon modes.
\end{abstract}

%\textcolor{red}{}

\section{The influence of hydration on the dynamics of the DNA.}

Tera-hertz waves have attracted considerable attention  owing to their increasing applications
and  possible biological effects. There is a need of careful studying the action of
the Tera-hertz waves on molecules of the DNA. The main point is whether they could generate elastic vibrations, or phonon modes, strong enough  
to damage a molecule of DNA.  Woolard et al, \cite{Woolard}, \cite{Woolard2},  have  reported 
the direct identification of the phonon modes of DNA with the help of Tera-hertz spectroscopy.  
Theoretical analysis of  the problem runs across difficulties because  the system's small size and the range  of hyper-sound frequencies 
thwart the use of the conventional mechanics of condensed media, \cite{fabelinsky}, \cite{Frenkel}. 
The problems  require  studying  the dynamics of the duplex of the DNA, the so-called inter-strand modes. 
The experimental research and the numerical modelling     
indicated that they are strongly attenuated, \cite{Urabe}, \cite{Tominaga},\cite{Lindsay},  \cite{prohofsky}. 
There are different opinions as to the extent of the damping. In papers \cite{Saxena_Zandt}, \cite{Saxena_Zandt_2}, \cite{Georghiou} 
the authors put forward arguments in favor the inter-strand modes not being over-damped.  
The recent experimental results by Woolard et al,  \cite{Woolard}, \cite{Woolard2}, indicate that the modes are strong enough to be
observed. 

The theoretical analysis of the elastic modes of the double helix has been performed, so far, within the framework of elasticity theory,
without taking into account viscous effects. In this paper we shall take into account the viscosity due to the hydration sheet, and 
deviations of the  structure of DNA from the perfect helical symmetry, see \cite{calladine}
Thus, we shall compare a molecule of DNA with a composite material, \cite{fung}.  

We rely on  Maxwell's model of viscoelastic dynamics.  According to the latter
one may cast the relation between stress, $F$, and strain,  $u$, in the form 
\begin{equation}
     \label{eq:maxwell}
     \frac{\partial F}{\partial t}  =  \lambda \frac{\partial u}{\partial t}  \; - \; \frac{1}{\tau} F 
\end{equation}
Here $\lambda$ is the elastic constant characteristic of a body, and $\tau$ is the characteristic time of relaxation.
The approach was employed for studying viscoelastic dynamics of liquids and amorphous bodies in the THz range of frequencies, \cite{Frenkel}.
It leads to important conclusions as to the mechanism of dissipation.
and  results in a theory that substantially differs from the conventional Navier-Stokes hydrodynamics.
The most spectacular evidence to the effect is the Mandelstam-Brillouin light scattering in liquids. 
The Navier-Stokes hydrodynamics gives the width of a line in the Mandelstam-Brilloiun  doublet larger than the width of the doublet itself, 
so that it should not be observed. In fact,  it is, \cite{fabelinsky}. According to the theory worked out 
by Leontovich and Mandelstam, \cite{fabelinsky}, the attenuation of sound ceases to depend on its frequency $\omega$ 
in the hyper-sound region,  whereas in the low frequency region, where the Navier-Stokes theory is valid, 
it varies as $\omega^2$.  To accommodate the hydration one shall recall that the   two nearest layers  form a kind of hydration water sheet,
the bulk of solvent being outside.  Chen and Prohofsky, \cite{chen} suggested that  the  water location  
have specific sites with respect to the double helix of DNA, which  form a kind of spine in the narrow groove. 
Thus, the hydration water  can be considered as an integral part of the helix,
essential to its stability  and dynamics, \cite{chen}.  Viscous effects depend on the exchange of water molecules between the hydration sheet
and the bulk of solvent, \cite{williams}.  They have a characteristic time of tens of ps.  In the spirit of the composite theory, \cite{fung}  , 
we may claim that the interplay of the elastic properties of molecules of DNA and the hydration result in a renormalization 
of elastic constants. Owing to the THz-range of frequencies we may use Maxwell's approach to viscoelasticity, \cite{Frenkel}.

\section{\label{sec:elastic}The elastic dynamics of torsional and
inter-strand modes}

Considering the dynamics of the DNA one has to take into account: \\
(1) the DNA having the two strands; \\
(2) the base-pairs being linked by the hydrogen bonds; \\
(3) the helical symmetry.

To estimate  the attenuation discussed above we employ  a one-dimensional lattice model 
of the DNA worked out in paper \cite{golo}.  The model accommodates these requirements, see the Appendix.
We  use the scheme worked out by
El Hasan and Calladine, \cite{calladine}, for the internal
geometry of the double helix of the DNA.  It describes the relative
position of one base with respect to the other in a Watson-Crick base-pair
and also the positions of the two base-pairs.  This is achieved by
introducing local frames for the bases and the base-pairs,
and translation-slides along their long axes.
We shall describe the relative positions of the
bases of a base-pair by  means of the vector $\vec Y$ directed along the
long axis ( $y-$  axis of paper \cite{calladine}), $\vec Y$ being equal to zero
when the base-pair is at equilibrium.
The relative positions of the base-pairs are described  by the torsional
angles $\phi_n$, which give deviations from the standard equilibrium twist of
the double helix.
Thus a twist of the DNA molecule, which does not involve inter-strand motion
or mutual displacements of the bases inside the pairs,
is determined by the torsional angles $\phi_n$ that are
the angles  of rotation of the base-pairs about the axis of the double-helix.
The  motions should correspond to the relative motion of
the bases inside the base-pairs.  
For each base-pair we have the reference frame in which
z-axis corresponds to the axis of the double helix,
y-axis to the long axis of the base-pair,
x-axis perpendicular to z- and y- axes (see Fig. 1 of paper \cite{calladine}).
At equilibrium the change in position of
adjacent base-pairs is determined only by the twist angle $\Omega$ of
the double helix.
We shall assume $\Omega = 2 \pi / 10$ as for the B-form of DNA.

The  above model provides a qualitative description  of the dynamics of the duplex. 
It indicates three modes of motion:
a torsional one that corresponds to the twist angle $\phi_n$, and two modes for the vector $\vec Y_n$
of inter-base displacements, \cite{golo}. Accommodating the relaxation effects in accord with the recipe given above
 (see the relevant calculations in the Appendix)
we may cast the dispersive equations in the form: \\
(1) for the modes  of the vectors $\vec Y_n$ we have
\begin{equation}
      \label{eq:vecY}
      \omega^2_{\alpha q} = \omega^2_Y \sin^2 \frac{\Omega + (-1)^{\alpha} aq}{2} 
      \; + \; \omega^2_s \; -  \;  \frac{1}{\tau} \omega_{\alpha q} , \qquad \alpha = 0,  1
\end{equation}
(2) for the twist mode of the duplex
\begin{equation}
      \label{eq:phi}
      \omega^2_q = \omega^2_t \, \sin^2\frac{aq}{2}  \; -  \;  \frac{1}{\tau}  \omega_q
\end{equation}
Here $\tau$  is the relaxation time, and
$\omega_Y, \; \omega_s, \omega_t$  are the characteristic frequencies for the motion of the $Y$, the stretching of hydrogen bonds, and
the twist of the duplex (see the Appendix). 
We see that the spectrum of the twist mode has a typical acoustic character,
whereas that for the $\vec Y_n$ has a local minimum determined by the helical angle, $\Omega$. Thus, the spectrum
of our model is in qualitative agreement with the conclusions of \cite{proh1}.  
Summarizing, we may state that there are: \\
(1) the internal degrees of freedom  which obey the dispersive law (\ref{eq:vecY}); \\ 
(2) the external ones generated by winding the strands round the axis of the double helix, and subject to equations  (\ref{eq:phi}).

In the range of large frequencies we may assume  $\omega \tau \gg 1$, so that  the approximate equation 
\begin{equation}
 \label{eq:disp2}
      k  \approx \frac{\omega}{c} \; - \; \frac{i}{2 \tau c}
 \end{equation}
 is true and the attenuation length, $\Lambda$, equals
 \begin{equation}
 \label{eq:disp3}
	  \Lambda  \approx 4 \pi  \tau c
 \end{equation}
 Here $c$ is the characteristic sound velocity of elastic excitations of a molecule of DNA.
 In fact, equation (\ref{eq:disp2}) is meaningful even for values $\omega \tau \ge 1$, for we obtain it with the formula of the Newtons binom 
 for exponent $1 /2$ and the imaginary correction of the third order.
 Following a rule of thumb, we may employ the last equation to find an estimate for the attenuation of elastic modes of DNA.

\section{Conclusions}
We see that the attenuation does not depend on frequency, if the latter is large enough, 
the length of attenuation being given by equations (\ref{eq:disp2}) and (\ref{eq:disp3}). 
Whether it could be physically significant depends on numerical values involved.
The velocity of elastic modes of the DNA is estimated as $\propto 10^5 cm/sec $. The relaxation time $\tau$ is determined by the exchange
of adsorbed water molecules and those of the bulk solvent. One may suppose  that the relaxation time is of the same order of magnitude,
or larger, as the residence time of adsorbed molecules, i.e   $10^{- \; 12} sec$, according to paper \cite{williams}. 
Phonons of a wave length of the order of $10 \; \AA$ have frequencies in the range of a few THz, and the condition $\omega \, \tau \gg 1$
is verified. At the same time we may infer from the equations given above that the attenuation length is of the order 
$\Lambda \propto 10^{- \; 6} cm $ or $100 \; \AA$, that  is of the order of the persistence length. 
Therefore, the relaxation effects are strong, but the above estimate leaves a room for the elastic modes of DNA being non-over-damped.

\section{Appendix}

For the convenience of the reader we shall recall the model, \cite{golo},
which relies on the earlier paper by H.Capellmann and W.Beim, \cite{biem}.
As was indicated above,we use the notations of the paper, \cite{calladine}.  
A twist of the DNA molecule, which does not involve inter-strand motion
or mutual displacements of the bases inside the pairs,
is determined by the torsional angles $\phi_n$ that are
the angles  of rotation of the base-pairs about the axis of the double-helix.
The  twist energy of the molecule is given by the equation
$$
     \sum_n\, \left[
		    \frac{I}{2} \, \dot{\phi}_n^2
		    +  \displaystyle{\frac{k}{2a^2}} \,
		       (\phi_{n+1} - \phi_n)^2
		 \right]
$$
in which $I$ is the moment of inertia, and $k$ is the twist
coefficients, which for the sake of simplicity
and taking into account the qualitative picture at which we aim,
are assumed to be   the same for all the base-pairs.
Inter-strand motions should correspond to the relative motion of
the bases inside the base-pairs, therefore
the kinetic energy due to this degree of freedom may
be cast in the form
$$
  \sum_n\, \frac{M}{2}\, \dot{\vec Y}_n^2
$$
where $M$ is the effective mass of a couple.

For each base-pair we have the reference frame in which
z-axis corresponds to the axis of the double helix,
y-axis to the long axis of the base-pair,
x-axis perpendicular to z- and y- axes (see Fig. 1 of paper \cite{calladine}).
At equilibrium the change in position of
adjacent base-pairs is determined only by the twist angle $\Omega$ of
the double helix.
We shall assume $\Omega = 2 \pi / 10$ as for the B-form of DNA.
To determine the energy due to the inter-strand displacements
we need to take into account the constraint
imposed by the helical structure of our system, which can be done as follows.
Let us confine ourself only to the torsional degrees of freedom of
the double lattice and assume the vectors $\vec Y_n$ being
parallel to x-y plane, or two-dimensional.
Consider the displacements $\vec Y_n,\, \vec Y_{n+1}$
determined within the frames of the two consecutive base-pairs, n, \, n+1.
Since we must compare the two vectors in the same frame,
we shall rotate  the vector
$\vec Y_{n+1}$ to the frame of the n-th base pair,
$$
  \vec Y^{\, back}_{n+1} =  R^{-1}(\phi)\, \vec Y_{n+1}
$$
Here $R^{-1}(\phi)$ is the inverse of a $2 \times 2$ matrix
of the rotation of the n-th frame to the (n+1)-one.  The argument is just the same as in the Kirchhoff theory 
of elastic rod.
It is important that the angle $\phi$ is given by  $   \Omega + \phi_{n+1} - \phi_n $.
Therefore, the energy due to the {\it inter-strand} stress reads
$$
    \sum_n \left\{
	       \frac{M}{2}\, \dot{\vec Y_n}^2
	       + \displaystyle{\frac{K}{2a^2}} \,
		  \left[ R^{-1}(\Omega + \phi_{n+1} - \phi_n)\, \vec Y_{n+1}
		    - \vec Y_n
		  \right]^2
	    \right \}
$$

At equilibrium the $\phi_n$ are equal to zero.
We suppose that the size of DNA molecule is small enough that
it can be visualized as a straight double helix, that is not larger than
the persistence length.  Hence the number of base-pairs,
$ N \le 150 $, approximately.
Combining the formulae given above we may write down the Lagrangian function of the system
 in the form
\begin{eqnarray}
  {\cal L} &=&
       \sum_n\, \left[
		    \frac{I}{2} \, \dot{\phi}_n^2
		    -  \displaystyle{\frac{\tau}{2a^2}} \,
		       (\phi_{n+1} - \phi_n)^2
		 \right] \nonumber \\
       &+& \sum_n \left\{
	       \frac{M}{2}\, \dot{\vec Y_n}^2
	       - \displaystyle{\frac{K}{2a^2}} \,
		  \left[ R^{-1}(\Omega + \phi_{n+1} - \phi_n)\, \vec Y_{n+1}
		    - \vec Y_n
		  \right]^2
	       + \frac{\epsilon}{2}\, \vec Y_n^2
	    \right \}    \label{eq:main}
\end{eqnarray}
in which  $K, k$ and $a$ are the torsional elastic constants
and the inter-pairs distance, correspondingly.
In summations given above
n is the number of a site corresponding to the n-th base-pair, and
$ n= 1,2, \ldots, N $, $N$ being the number of pairs in the segment of
the DNA under consideration. The terms, $\epsilon / 2 \, \vec Y^2$
accommodate  the energy of the inter-strand {\it separation}
due to the {\it slides of the bases inside the base-pairs}.
It should be noted that the dynamical  variables $\phi_n$ and $\vec Y_n$
are of the same order of magnitude, that is the first.

We shall introduce the relaxation following  Maxwell's prescription  (\ref{eq:maxwell}).
We may cast (\ref{eq:maxwell}) into the explicit dependence of the force on the displacement
$$ 
  F = \lambda u \; - \; \left[ \lambda ( 1 + \tau \frac{\partial}{\partial t} )^{- \; 1}  
                        \right] u
$$
For very large relaxation times we get the Hooke law, $ F = \lambda u$, and for very small relaxation times, 
that is very large dissipation, the external force is extinguished, $ F = 0 $. Thus, the relaxation amounts 
to changing scalar elastic constants $\lambda$ for the operator ones, \cite{Frenkel},
$$
  \lambda \rightarrow \hat{\lambda} \; 
   = \; \lambda \left( 1 \; - \; {\cal F}  \right)  
$$
Let us consider the general situation which fully describes 
the problem at hand. We have a system given  by a Lagrangian function 
$     {\cal L} =  {\cal L} (u_1, u_2, \ldots, u_n; \dot{u}_1, \dot{u}_2, \ldots, \dot{u}_n;) $  
In our case it is the function determined by the equations for the kinetic and the potential energy given above.
The equations of motion have the Lagrangian form
$$
   \frac{d}{d t} \frac{\cal L}{\dot{u}_k}  = \frac{\partial {\cal L}}{\partial u_k}
$$
In accord with Maxwell's  prescription we set
$$
  \frac{d F_k}{d t} = \frac{\partial {\cal L}}{\partial \dot{u}_k} \; - \; \frac{1}{\tau} F_k, 
  \qquad \mbox{or}  \qquad
  F_k = \left( 1 \; + \; \tau \frac{d}{d t}  \right)^{- \; 1} \frac{\partial {\cal L}}{\partial \dot{u}_k}
$$
On  substituting the expression for the force in the right hand sides of the Lagrangian equations, and obtain the equation
$$
      \frac{d}{d t} \frac{\cal L}{\dot{u}_k}  = - \; \frac{1}{\tau} \frac{\partial {\cal L}}{\partial \dot{u}_k}  
                                                \; + \; \frac{\partial {\cal L}}{\partial u_k}
$$
from which we infer equations (\ref{eq:disp2}),(\ref{eq:disp3}).

\end{document}